# Strong Electric Field in 2D Graphene: The Integer Quantum Hall Regime from a Different (Many-Body) Perspective

## Georgios Konstantinou[1], Konstantinos Moulopoulos[2]

[1]FOSS Research Centre for Sustainable Energy, PV Technology, University of Cyprus, Nicosia, Cyprus
[2]Department of Physics, University of Cyprus, Nicosia, Cyprus
Email: ph06kg1@ucy.ac.cy, cos@ucy.ac.cy

## Abstract

We investigate the emerging consequences of an applied strong in-plane electric field on a macroscopically large graphene sheet subjected to a perpendicular magnetic field, by determining in exact analytical form various many-body thermodynamic properties and the Hall coefficient. The results suggest exotic possibilities that necessitate very careful experimental investigation. In this alternate form of Quantum Hall Effect, non-linear phenomena related to the global magnetization, energy and Hall conductivity (the latter depending on the strengths of magnetic B- and electric E-fields) emerge without using perturbation methods, to all orders of E-field and B-field strengths. Interestingly enough, when the value of the electric field is sufficiently strong, fractional quantization also emerges, whose topological stability has to be verified.

## Keywords

Graphene, Landau Levels, Strong Electric Field Effects, Hall Conductivity, Magnetization, Quantum Hall Effect, Thermodynamics

## 1. Introduction

Dirac-type materials, such as Topological Insulators, monolayer graphene, and three-dimensional (3D) Dirac and Weyl semimetals, appear nowadays as stable (actually very robust) topological phases of matter, displaying behavioral patterns that produce new physics at a very fundamental level and at the same time give the possibility of exotic future applications [1] [2] [3]. What make their fundamental properties so fascinating are the well-known dissipationless surface states that can propagate without any resistance and give rise to nontrivial topo-logical properties that are currently under intense investigation. When certain types of such materials (*i.e.* 2D Topological Insulators) are subject to a perpendicular magnetic field, they may as well undergo a phase transition to Quantum Hall Insulators [4], violating the time reversal symmetry that controls the topology of the surface states. Normally, there is a transverse (to B) small electric field E, which—to first order in E—is responsible for the macroscopic quantization of the Hall conductivity [5], and which is the central quantity in the present paper. Interestingly enough, the strong E-field regime has not been investigated in sufficient detail so far, in particular with respect to the role of the E-field on thermodynamic many-body properties (see however [6], and for some earlier attempts see [7]-[14]), as these properties are determined in the noninteracting electrons framework (the one that, in any case, pertains to the Integer Hall Effect regime). In this work, we present potential consequences (on thermodynamic and transport properties) of a strong electric field applied tangentially to a macroscopic 2D graphene sheet, when also subjected to a perpendicular magnetic field of arbitrary strength.

Let us start with the graphene energy spectrum when a monolayer is subjected to an in-plane electric $E$ (taken along the $x$-direction) and a perpendicular magnetic field B in the z-direction (and let us focus on the positive branch, and also ignore the Zeeman interaction term), and take the Landau gauge $\boldsymbol{A} = (0, xB, 0)$ in which the energy spectrum turns out to be (through a Dirac equation procedure similar to the one in [15])

$$\varepsilon_{n,k_y} = \sqrt{2n\hbar eBu_f}\left(1-\beta^2\right)^{3/4} + \hbar u_f \beta k_y \,, \tag{1.1}$$

with $\beta = E/u_f B$ a dimensionless parameter (always supposed to be lower than unity, $\beta < 1$), $n = 0, 1, 2, 3$. The Landau Level index for the positive branch, $u_f$ is the Fermi velocity, and $k_y$ is the wave vector along the $y$-direction. We also find that the guiding center operator's eigenvalue (projected on the $x$-axis) $X_0$ reads

$$X_0 = l_B^2 k_y - \text{sgn}\left(n\right)\frac{\beta l_B \sqrt{2|n|}}{\left(1-\beta^2\right)^{1/4}} \,, \tag{1.2}$$

with $l_B = \sqrt{\hbar c/eB}$ the magnetic length and $\text{sgn}\left(n\right)$ is the sign function. Due to the spatial confinement in the x-direction, the guiding center operator $X_0$ may acquire any value in the following range:

$$-\frac{L_x}{2} \leq X_0 \leq \frac{L_x}{2} \tag{1.3}$$

with $L_x$ being the x-direction size of the system, which is here supposed to be macroscopically large. Each Landau Level, defined by different values of index n, contains $\Phi/\Phi_0$ independent quantum states, with $\Phi = BS$ the magnetic flux penetrating the 2D graphene sheet ($S$ is the area) and $\Phi_0 = hc/e$ is the flux quantum. Now, because of the spin and valley degeneracies that are present in graphene, each Landau Level (L.L.) may accommodate up to $4\Phi/\Phi_0$ spinful electrons, according to the Pauli exclusion principle excluding the lowest level $n = 0$, which can accommodate only up to $2\Phi/\Phi_0$ spinful electrons, due to mutual sharing with the holes.

As a next step it is convenient (for the thermodynamic calculations followed below), to express (1.1) as a function of the guiding center operator, namely

$$\varepsilon_{n,X_0} = \frac{\sqrt{2n\hbar eBu_f}}{\left(1-\beta^2\right)^{1/4}} + eEX_0 \, . \tag{1.4}$$

A few remarks are then in order about the energy gap (the inter-L.L. gap, for a constant guiding center) determined by:

$$\delta\varepsilon_n = \frac{\sqrt{2\hbar eBu_f}}{\left(1-\beta^2\right)^{1/4}}\left(\sqrt{n+1}-\sqrt{n}\right). \tag{1.5}$$

Unlike conventional semiconductors, the energy gap has an $E$-field dependence. As can easily be seen from (1.5), the larger the $E$-field gets, the larger the energy gap becomes. On the other hand, the larger the L.L. index, the smaller the energy gap. This interplay will play a major role later on, when we consider the thermodynamic occupations of the energy levels. One can always prove that, for a given L.L. index and a value of $E$-field (such that $\beta < 1$) there will always be an unavoidable overlap (states of greater $n$ values have lower energy than states with lower $n$ values). We can set conditions (for arbitrary $E$ and $n$) for which this kind of overlap is avoided as:

$$\varepsilon_{n,L_x/2} \le \varepsilon_{n+1,-L_x/2} \tag{1.6}$$

To picture this more properly, we provide examples in the graphs shown in **Figure 1**.

Condition (1.6) along with (1.4) results in:

$$\frac{\sqrt{2\hbar eBu_f}}{\left(1-\beta^2\right)^{1/4}}\left(\sqrt{n+1}-\sqrt{n}\right) \ge eEL_x \quad \text{or} \quad \delta\varepsilon_n \ge eEL_x \tag{1.7}$$

That is, in words, when the work performed by the electric field is smaller than the energy gap at a certain $X_0$, no overlap is observed between the L.Ls $n$ and $n + 1$. Generally, for strong enough electric field or for a small enough L.L. index, the above inequality will be true (see **Figure 2**); but as the L.L. index of occupied levels gets larger (hence for a large number of electrons $N$) the energy gaps between adjacent L.Ls will become lower, until an inevitable gap closing occurs (**Figure 2** providing a concrete example).

We define $n = i_F$ to be the topmost L.L., which for a given value of $E$ and $B$ maintains an energy gap with its adjacent L.Ls: ($n = i_F - 1$, $n = i_F + 1$) Clearly, L.L index $n = i_F + 1$ no longer separates from $i_F + 2$ with an energy gap, and in this case, inequality (1.7) reverses direction: $\delta\varepsilon_{i_F+1} \le eEL_x$. In other words, L.Ls with $n = 0,1,2,\cdots,i_F$ do not overlap, while L.Ls with higher quantum numbers ($n = i_F, i_F+1, i_F+2, \cdots$) do overlap.

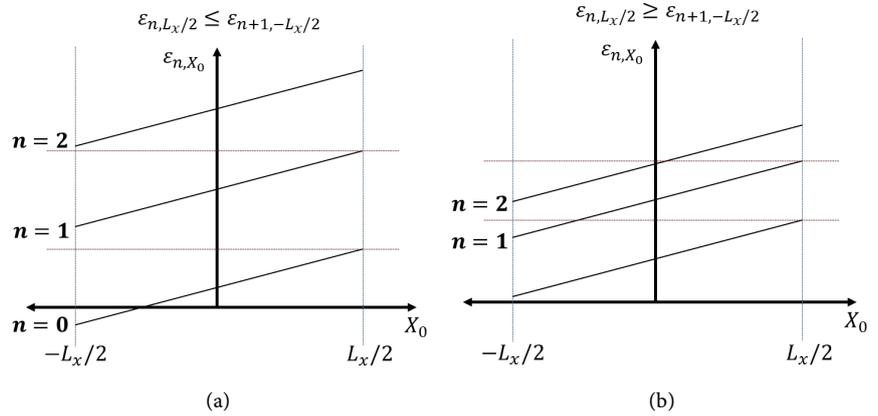

**Figure 1.** (a) An example of an $E$-field value (strong) that do not cause overlaps between different L.Ls; (b) Another example of a weaker $E$-field that causes overlaps between adjacent L.Ls.

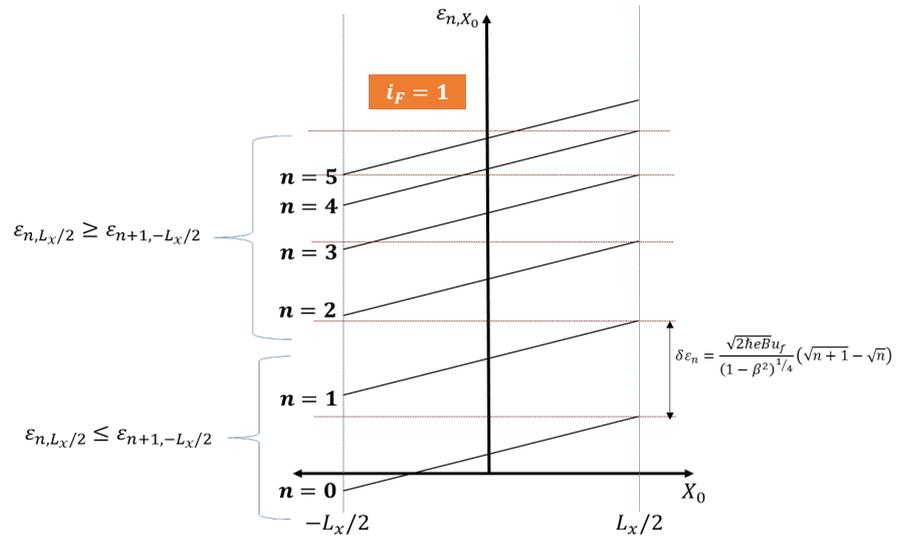

**Figure 2.** In an arbitrary, fixed $E$-field strength, overlaps occur as L.L. index gets larger. The overlaps are indicated by the conditions (inequalities) shown at the left of the figure. The red lines indicate the top-most energetic state in a given L.L., in comparison with the next adjacent L.L. lower state. The stronger the $E$-field gets, the larger the energy gap becomes, and the overlap will occur at an energetically high L.L. In the above example, overlaps occur between ($n = 2$, $n = 3$), ($n = 3$, $n = 4$), ($n = 4$, $n = 5$, $n = 6$ (not shown)). Levels $n = 0$, $n = 1$ do not overlap and they provide independent energy states when following an occupation procedure. In this case, $i_F = 1$.

The above is a generic case for an arbitrary value of $E$ and $n$. Of course there might be cases where overlaps start from $n = 0$ (for a low enough $E$-field), in which case $i_F = 0$, and all L.L.s with $n > 0$ overlap. In this case, with no energy gap present at all, graphene will gain a metallic character. Equation (1.7) with input $n = i_F + 1$ becomes then:

$$\frac{\sqrt{2\hbar e B}u_f}{\left(1 - \beta^2\right)^{1/4}}\left(\sqrt{i_F + 2} - \sqrt{i_F + 1}\right) \leq eEL_x \tag{1.8}$$

## 2. The Strong *E*-Field Regime

After the above discussion and definitions, we proceed to thermodynamic occupations of the graphene's energy levels at zero temperature. For this purpose, we consider a collection of $N$ electrons at $T = 0$, which fill the lowest energy levels until the Fermi energy denoted by $\varepsilon_F$. In reality, the Fermi energy is not constant when there is an electric field running through the system; what we then mean by Fermi energy is actually a Fermi point, which is the topmost occupied state in the energy diagram. We also make the supposition that this Fermi point is located on a L.L. indexed with $n = \rho - 1$ ( $\rho = 1, 2, 3, \cdots$ ), so that there are always $\rho$ L.Ls occupied at any time (the last level $n = \rho - 1$ being generally partially occupied). First, we will focus on the special case where all $\rho$ Landau Levels are not overlapping, and can be occupied independently by the $N$ electrons. In this case the following relation must be satisfied:

$$\rho - 1 \le i_F \tag{2.1}$$

and for strong enough magnetic fields, it is guaranteed that Equation (2.1) will always be satisfied, and no overlap between L.Ls with different quantum numbers will be observed. In what follows, we will consider a constant, strong $E$-field, while the magnetic field may vary, but always in a way that satisfies Equation (2.1).

Considering that the L.L. with $n = 0$ only has a capacity for $2\Phi/\Phi_0$ electrons, and that all the other $n \ne 0$ L.Ls may host up to $4\Phi/\Phi_0$ electrons, we find that in order to have $\rho$ L.Ls occupied, the following inequality must hold:

$$(2\rho - 3)\frac{2\Phi}{\Phi_0} \le N \le (2\rho - 1)\frac{2\Phi}{\Phi_0} \tag{2.2}$$

Note that when $\rho = 1$ then $0 \le N \le \dfrac{2\Phi}{\Phi_0}$, as $N$ is always a positive number.

Treating $N$ as a constant, we can solve (2.2) with respect to magnetic field $B$:

$$\frac{1}{2(2\rho - 3)} n_A \Phi_0 \ge B \ge \frac{1}{2(2\rho - 1)} n_A \Phi_0 \tag{2.3}$$

with $n_A$ the electronic surface density and $\Phi_0$ the flux quantum. Also note that we have considered the special case where $\rho \le i_F + 1$, where all L.Ls are well separated with an energy gap, and no inter-L.L. overlapping occurs. (The special case of nonzero overlaps will be considered in the next Section). Note that Equation (2.3) can also describe the well-known unconventional Quantum Hall Effect in graphene, with the Hall conductivity given by the well-known relation:

$$\sigma_H = \frac{en_A}{B} = \left(\rho - \frac{1}{2}\right)\frac{4e^2}{h} \tag{2.4}$$

If one replaces $B$ by the value $B = \dfrac{1}{2(2\rho - 1)} n_A \Phi_0$; $\sigma_H$ is then quantized in half-integer multiples of $\dfrac{4e^2}{h}$.

In the case we are considering the total energy of the system (minimized at $T$

= 0) is given as a sum over all states lower than "the Fermi state" or Fermi point, namely

$$E = \sum_{n,X_0} \varepsilon_{n,X_0} = \sum_{n,X_0} \frac{\sqrt{2n\hbar eB}u_f}{\left(1-\beta^2\right)^{1/4}} + eEX_0 \qquad (2.5)$$

In the thermodynamic limit $L_x \to \infty$, we may approximate the sum with respect to $X_0$ as follows:

$$\sum_{X_0} \to 4\frac{BL_y}{\Phi_0}\int_{-\frac{L_x}{2}}^{\frac{L_x}{2}}dX_0 \quad \text{for} \quad n>0 \quad \text{and} \quad \sum_{X_0} \to 2\frac{BL_y}{\Phi_0}\int_{-\frac{L_x}{2}}^{\frac{L_x}{2}}dX_0 \quad \text{for} \quad n=0 . \qquad (2.6)$$

The total internal energy of the system can then be separated in the energy of the fully occupied bands plus the energy of the partially occupied last L.L. (that contains the Fermi point, $n = \rho - 1$):

$$E = E_{\text{full}} + E_{\text{part}} \qquad (2.7)$$

In what follows, we will consider the case $\rho > 1$, where the contribution of the lowest L.L. $n = 0$ is negligible. Given that all the bands up to $n = \rho - 2$ are fully occupied, we may write:

$$E_{\text{full}} = 4\frac{BL_y}{\Phi_0}\sum_{i=1}^{\rho-2}\int_{-\frac{L_x}{2}}^{\frac{L_x}{2}}dX_0\frac{\sqrt{2n\hbar eB}u_f}{\left(1-\beta^2\right)^{1/4}} + 4\frac{BL_y eE}{\Phi_0}\sum_{i=1}^{\rho-2}\int_{-\frac{L_x}{2}}^{\frac{L_x}{2}}X_0 dX_0$$
$$= 4\frac{\Phi}{\Phi_0}\frac{\sqrt{2\hbar eB}u_f}{\left(1-\beta^2\right)^{1/4}}\sum_{n=1}^{\rho-2}\sqrt{n} \qquad (2.8)$$

To determine $E_{\text{part}}$, we must first determine the $X_0$ value at the Fermi point (limit of the $X_0$ integration at the $n = \rho - 1$ L.L.). From Equation (1.2) we get for $k_y = 2\pi l/L_y$ and $n = \rho - 1$:

$$X_0 = l_B^2\frac{2\pi l}{L_y} - \frac{\beta l_B\sqrt{2(\rho-1)}}{\left(1-\beta^2\right)^{1/4}} \qquad (2.9)$$

The quantum number $l$, appearing in (2.9) has a starting value $l_0$ that can be determined by setting $X_0 = -\frac{L_x}{2}$ as follows:

$$l_0 = -\frac{\Phi}{2\Phi_0} + \frac{L_y}{2\pi l_B}\frac{\beta\sqrt{2(\rho-1)}}{\left(1-\beta^2\right)^{1/4}} \qquad (2.10)$$

Now, using the fact that the last L.L. contains $N - (2\rho-3)\frac{2\Phi}{\Phi_0}$ electrons, we have all the necessary information to determine the guiding center value of the electron placed at the Fermi point:

$$X_{0,F} = \frac{2\pi}{L_y}l_B^2\left[\frac{N}{4} - (\rho-1)\frac{\Phi}{\Phi_0}\right] \qquad (2.11)$$

For example, when $N = 2(2\rho - 3)\dfrac{\Phi}{\Phi_0}$, (meaning that the last L.L. is empty of electrons), we get from (2.11):

$$X_{0,F} = \frac{2\pi}{L_y} l_B^2 \left[ \frac{2(2\rho - 3)}{4} - (\rho - 1) \right] \frac{\Phi}{\Phi_0} = -\frac{\pi}{L_y} l_B^2 \frac{\Phi}{\Phi_0} = -L_x/2, \qquad (2.12)$$

whereas for $N = 2(2\rho - 1)\dfrac{\Phi}{\Phi_0}$ we have that $X_{0,F} = L_x/2$, explicitly demonstrating the correctness of our results. The remaining task to carry out is to calculate the energy of the partially occupied L.L. $n = \rho - 1$

$$E_{part} = 4\frac{BL_y}{\Phi_0} \int_{\frac{L_x}{2}}^{X_{0,F}} dX_0 \frac{\sqrt{2(\rho - 1)\hbar e B u_f}}{(1 - \beta^2)^{1/4}} + eE4\frac{BL_y}{\Phi_0} \int_{\frac{L_x}{2}}^{X_{0,F}} X_0 dX_0$$

$$= 4\frac{BL_y}{\Phi_0} \frac{\sqrt{2(\rho - 1)\hbar e B u_f}}{(1 - \beta^2)^{1/4}} \left( X_{0,F} + \frac{L_x}{2} \right) + eE4\frac{BL_y}{\Phi_0} \left( \frac{X_{0,F}^2}{2} - \frac{L_x^2}{8} \right) \qquad (2.13)$$

Substituting Equation (2.11) into (2.13) we obtain (in units of Fermi energy (in the absence of the electromagnetic field) $\varepsilon_F = \hbar u_F \sqrt{\pi n_A}$, per particle)

$$\frac{E_{part}}{N} = \frac{\varepsilon_F}{(1 - \beta^2)^{1/4}} \left( \left( \frac{B}{n_A \Phi_0} \right)^{1/2} \sqrt{4(\rho - 1)} - 2\sqrt{4(\rho - 1)} \left( \frac{B}{n_A \Phi_0} \right)^{3/2} [2\rho - 3] \right)$$

$$+ \left( \frac{NE}{8L_y} \frac{h}{B} - \frac{eSE}{L_y}(\rho - 1) + \frac{eEBL_x}{2n_A \Phi_0} [4\rho^2 - 8\rho + 3] \right) \qquad (2.14)$$

Finally, adding Equations (2.14) and (2.8) we arrive at the following result:

$$\frac{E}{N} = \frac{\varepsilon_F}{(1 - \beta^2)^{1/4}} \left( 4 \left( \frac{B}{n_A \Phi_0} \right)^{3/2} \left[ 2\sum_{n=1}^{\rho-2} \sqrt{n} - [2\rho - 3]\sqrt{(\rho - 1)} \right] \right.$$

$$\left. + \left( \frac{B}{n_A \Phi_0} \right)^{1/2} \sqrt{4(\rho - 1)} \right) + \left( \frac{NE}{8L_y} \frac{h}{B} - \frac{eSE}{L_y}(\rho - 1) + \frac{eEBL_x}{2n_A \Phi_0} [4\rho^2 - 8\rho + 3] \right) \qquad (2.15)$$

with $\sum_{n=1}^{\rho-2} \sqrt{n} = -\zeta\left( -\frac{1}{2}, \rho - 1 \right) - \dfrac{\zeta\left( \frac{3}{2} \right)}{4\pi}$ where $\zeta\left( -\frac{1}{2}, \rho - 1 \right)$ is the Hurwitz Zeta function and $\zeta\left( \frac{3}{2} \right)$ is the Riemman function. Note the interesting fact that, terms proportional to electric field strength $E$ result in the following magnetization:

$$\frac{M_E}{N} = -\frac{\partial \frac{E}{N}}{\partial B} = \frac{NE}{8L_y} \frac{h}{B^2} - \frac{eEL_x}{2n_A \Phi_0} [4\rho^2 - 8\rho + 3], \qquad (2.16)$$

which, when considered at full band occupation (meaning that $B = \dfrac{1}{2(2\rho - 1)} n_A \Phi_0$) yields:

$$\frac{M_E}{N} = \frac{NE}{8L_y}\frac{4h(2\rho-1)^2}{n_A^2\Phi_0^2} - \frac{eEL_x}{2n_A\Phi_0}\left[4\rho^2 - 8\rho + 3\right]$$

$$= \frac{EL_x}{n_A}\left(\rho - \frac{1}{2}\right)\frac{2e^2}{h} = \frac{L_x}{n_A}\frac{\sigma_H}{2}E \tag{2.17}$$

*i.e.* the proportional constant is equal to half Hall conductivity, similar to the corresponding result in a conventional semiconductor case [3]. (Plots of the field-free Energy and Magnetization are given in **Figure 3**).

# 3. The Weak *E*-Field Regime

We now proceed to a considerably more difficult case: the low *E*-field regime. In this regime, as the electric field becomes weaker, the energy gap gets smaller. As a result, there will be an unavoidable point where some L.Ls will overlap (**Figure 1(b)**), and occupational patterns turn out to be more complex. Inequality (1.7) is no longer true for all *n*'s (it will indeed be true only for L.Ls with quantum number $\leq i_F$). If, for simplicity reasons, we suppose for a moment that the energy spectrum configuration stays the same as before, and that the only change we have is a larger number of electrons *N* to be placed on the available states, things become a bit clearer. L.Ls up to $i_F$ ($n = 0,1,2,\cdots,i_F$) won't overlap with any of the rest L.Ls, while L.Ls with $n = i_F, i_F+1, i_F+2, \cdots, \rho-1$ will indeed overlap. In this case, we may also have that $\rho-1 \geq i_F$. Recall from the previous Section that in addition, the following relation must also hold: $\beta < 1$.

We fix the Fermi point at $n = \rho-1$, with energy given by:

$$\varepsilon_F = \frac{\sqrt{2(\rho-1)\hbar eBu_f}}{(1-\beta^2)^{1/4}} + eEX_{0F} \tag{3.1}$$

with $X_{0F}$ the guiding center position of the last energetically highest electron, occupying the Fermi point state. For the example case shown in **Figure 4**, and further supposing that the Fermi point is located at $X_{0F} = -L_x/2$ (**Figure 5**) for simplicity reasons, we have that:

$$\varepsilon_F = \frac{\sqrt{8\hbar eBu_f}}{(1-\beta^2)^{1/4}} - eE\frac{L_x}{2} \tag{3.2}$$

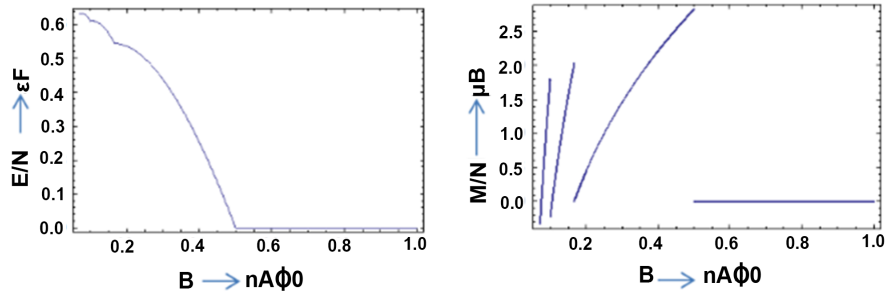

**Figure 3.** Energy in units of Fermi energy $\varepsilon_F$ and magnetization in units of Bohr magneton $\mu_B$ as functions of the magnetic field when the electric field is switched off.

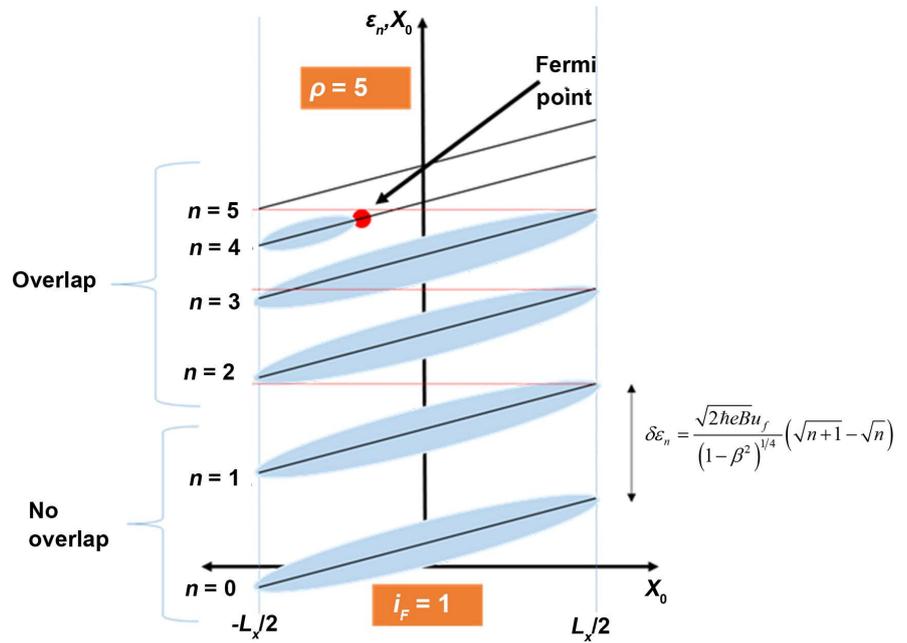

**Figure 4.** Diagram showing the occupations (navy blue) of the lowest energy states up to the Fermi point (red dot). Note that this is the optimal energy configuration for $i_F = 1$ and $\rho = 5$. Landau levels indexed with $n = 0$ and $n = 1$ do not overlap, while $n = 2$ overlaps with $n = 3$, and $n = 3$ overlaps with $n = 4$ and $n = 5$ (currently not occupied). If the Fermi point were located at *i.e.* $n = 1$, meaning that $\rho = 2$, we would then expect that the Hall conductivity would be quantized according to Equation (2.4). But, now, because of the inclusion of the overlaps, there is no need for an integer quantization, as we shall see in the main text.

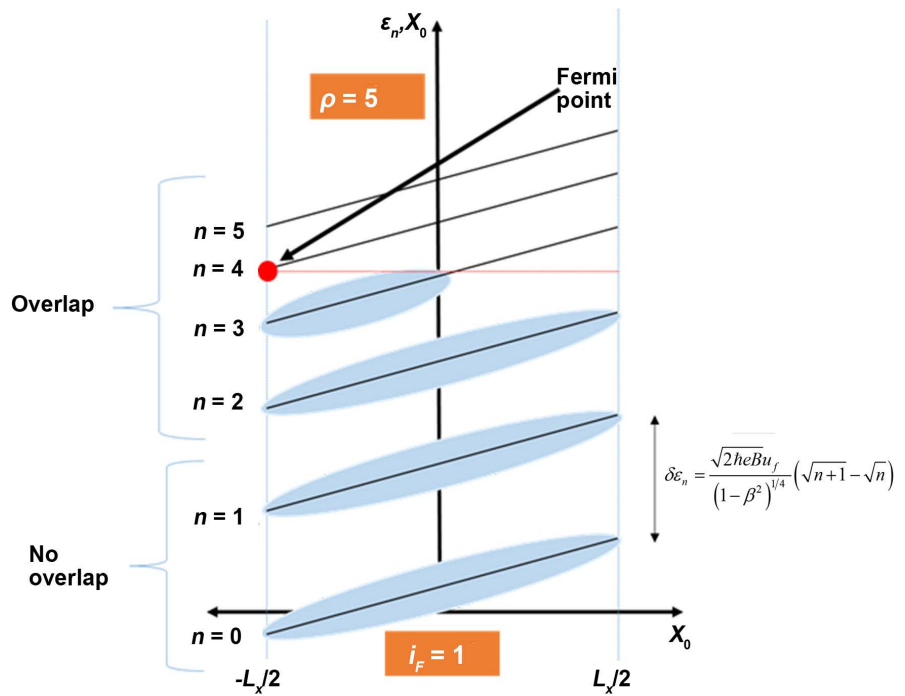

**Figure 5.** An example of the position of Fermi point for the purpose of the calculation in the main text.

Now, from the **Figure 5**, we observe that in this configuration, L.Ls $n = 0$, 1 and 2 are fully occupied, while L.L. $n = 3$ is only partially occupied. To determine the exact number of states occupied in L.L. $n = 3$ we examine its intersection with the Fermi point (red line in **Figure 5**):

$$\frac{\sqrt{6\hbar eB}u_f}{\left(1-\beta^2\right)^{1/4}} + eEX_{0F} = \frac{\sqrt{8\hbar eB}u_f}{\left(1-\beta^2\right)^{1/4}} - eE\frac{L_x}{2} \tag{3.3}$$

which yields for $X_{0F}$:

$$X_{0F} = -\frac{L_x}{2} + \frac{\sqrt{\hbar eB}u_f}{eE\left(1-\beta^2\right)^{1/4}}\left(\sqrt{8}-\sqrt{6}\right) \tag{3.4}$$

From the above, we may determine the initial and final values of the $k_x$ (that is, $l_0$ and $l_f$), namely

$$l_0 = -\frac{\Phi}{2\Phi_0} + \frac{L_y}{2\pi l_B}\frac{\sqrt{6}\beta}{\left(1-\beta^2\right)^{1/4}} \tag{3.5}$$

$$l_f = -\frac{\Phi}{2\Phi_0} + \frac{BL_y}{Eh}\frac{\sqrt{\hbar eB}u_F}{\left(1-\beta^2\right)^{1/4}}\left(\sqrt{8}-\sqrt{6}\right) + \frac{EL_y}{hu_F}\frac{e\hbar\sqrt{6}}{\sqrt{eB}\left(1-\beta^2\right)^{1/4}} \tag{3.6}$$

The number of states in L.L. $n = 3$ is then given by the difference of the initial and final values of $l$, namely

$$l_f - l_0 = \frac{BL_y}{Eh}\frac{\sqrt{\hbar eB}u_F}{\left(1-\beta^2\right)^{1/4}}\left(\sqrt{8}-\sqrt{6}\right) \tag{3.7}$$

and the total number of states below the Fermi point reads:

$$g = \frac{\Phi}{\Phi_0}\bigg|_{n=0} + 2\frac{\Phi}{\Phi_0}\bigg|_{n=1,2} + \frac{BL_y}{Eh}\frac{\sqrt{\hbar eB}u_F}{\left(1-\beta^2\right)^{1/4}}\left(\sqrt{8}-\sqrt{6}\right) \tag{3.8}$$

Because all these states are filled with electrons (4 electrons in each state for $n > 0$ and 2 electrons for $n = 0$) we have that:

$$\begin{aligned} N &= 2\frac{\Phi}{\Phi_0} + 8\frac{\Phi}{\Phi_0} + 4\frac{BL_y}{Eh}\frac{\sqrt{\hbar eB}u_F}{\left(1-\beta^2\right)^{1/4}}\left(\sqrt{8}-\sqrt{6}\right) \\ &= 10\frac{\Phi}{\Phi_0} + 4\frac{BL_y}{Eh}\frac{\sqrt{\hbar eB}u_F}{\left(1-\beta^2\right)^{1/4}}\left(\sqrt{8}-\sqrt{6}\right) \end{aligned} \tag{3.9}$$

Equation (3.9) yields for the Hall conductivity:

$$\sigma_H = \frac{en_A}{B} = 10\frac{e^2}{h} + 4\frac{eL_y}{SEh}\frac{\sqrt{\hbar eB}u_F}{\left(1-\beta^2\right)^{1/4}}\left(\sqrt{8}-\sqrt{6}\right) \tag{3.10}$$

Although (3.10) does not necessarily imply that irrational quantization in graphene is possible, it is quite interesting to notice how the integer quantization (the first term) is destroyed as long as L.Ls become intermixed due to the extra overlaps—which in turn are induced by the $E$-field. The least one can gain out of these calculations is the importance of the energy gap, and the possible forma-tion of localized and extended states that lie inside this gap. When the gap is destroyed, the system becomes metallic; it may thus not be unreasonable to find a Hall conductivity that is electric/magnetic field-dependent, destroying the plateaux formation. This is something that needs to be further investigated experimentally.

## 4. Conclusion

In this work, a thermodynamic study has been conducted with respect to a 2D Graphene monolayer subjected to crossed electric and magnetic fields. Thermodynamic quantities like the global energy and magnetization have been exactly and analytically determined, in combination with transport properties, *i.e.* the Hall conductivity. The results suggest exotic possibilities that are here pointed out as a natural outcome of an exact and careful calculation in the noninteracting electrons many-body framework, and these necessitate exceedingly careful experimental investigation. Finally, with respect to the range of validity, although our results involve no approximations whatsoever, the general role of disorder in combination with the above physics of the strong fields is certainly something that needs further study.

## References

[1] Franz, M. and Molenkamp, L., Eds. (2013) Contemporary Concepts of Condensed Matter Science. Volume 6. Topological Insulators. Elsevier, Amsterdam.

[2] Hideo Aoki, S. and Dresselhaus, M., Eds. (2014) Physics of Graphene. Springer, Berlin. https://doi.org/10.1007/978-3-319-02633-6

[3] Takahashi, R. (2015) Topological States on Interfaces Protected by Symmetry. Chapter 4: Weyl Semimetal in a Thin Topological Insulator. Springer, Berlin. https://doi.org/10.1007/978-4-431-55534-6

[4] Prange, R.E. and Girvin, S.M. (1990) The Quantum Hall Effect. Springer, Berlin. https://doi.org/10.1007/978-1-4612-3350-3

[5] Yoshioka, D. (2002) The Quantum Hall Effect. Springer, Berlin. https://doi.org/10.1007/978-3-662-05016-3

[6] Konstantinou, G. and Moulopoulos, K. (2017) Ground State Thermodynamic and Response Properties of Electron Gas in a Strong Magnetic and Electric Field: Exact Analytical Solutions for a Conventional Semiconductor and for Graphene. *Journal of Applied Mathematics and Physics*, **5**, Article ID: 74872. https://doi.org/10.4236/jamp.2017.53055

[7] Lukose, V., Shankar, R. and Baskaran, G. (2007) Novel Electric Field Effects on Landau Levels in Graphene. *Physical Review Letters*, **98**, Article ID: 116802. https://doi.org/10.1103/PhysRevLett.98.116802

[8] Kawaji, S., Hirakawa, K. and Nagata, M. (1993) Device-Width Dependence of Plateau Width in Quantum Hall States. *Physica B: Condensed Matte*, **184**, 17. https://doi.org/10.1016/0921-4526(93)90313-U

[9] Nachtwei, G. (1999) Breakdown of the Quantum Hall Effect. *Physica E: Low-Dimensional Systems and Nanostructures*, **4**, 79. https://doi.org/10.1016/S1386-9477(98)00251-3

[10] Ebert, G., *et al.* (1983) Two-Dimensional Magneto-Quantum Transport in GaAs/AlGaAs

Heterostructures under Non-Ohmic Conditions. *Journal of Physics C*, **16**, 5441. https://doi.org/10.1088/0022-3719/16/28/012

[11] Lilly, M.P., Cooper, K.B., Eisenstein, J.P., Pfeiffer, L.N. and West, K.W. (1999) Evidence of an Anisotropic State of Two-Dimensional Electrons in High Landau Levels. *Physical Review Letters*, **82**, 394-397. https://doi.org/10.1103/PhysRevLett.82.394

[12] Kazarinov, R.F. and Luryi, S. (1982) Quantum Percolation and Quantization of Hall Resistance in Two-Dimensional Electron Gas. *Physical Review B*, **25**, 7626-7630. https://doi.org/10.1103/PhysRevB.25.7626

[13] Tsemekhman, V., *et al.* (1997) Theory of the Breakdown of the Quantum Hall Effect. *Physical Review B*, **55**, R10201. https://doi.org/10.1103/PhysRevB.55.R10201

[14] Kramer, T. (2006) Landau Level Broadening without Disorder, Non-Integer Plateaus without Interactions—An Alternative Model of the Quantum Hall Effect. *Revista Mexicana de Fisica*, **52**, 49-55.

[15] Peres, N.M.R. and Castro, E.V. (2007) Algebraic Solution of a Graphene Layer in a Transverse Electric and Perpendicular Magnetic Fields. *Journal of Physics: Condensed Matter*, **19**, Article ID: 406231. https://doi.org/10.1088/0953-8984/19/40/406231